\documentclass[final,numberedheadings]{aipproc}
\usepackage{amsmath,amssymb,mathrsfs}
\usepackage{wrapfig}
\usepackage{ccaption}
\usepackage[active]{srcltx}
\usepackage[matrix,frame,arrow]{xy}
\usepackage{amsmath}

\newcommand{\qw}[1][-1]{\ar @{-} [0,#1]}
\newcommand{\qwx}[1][-1]{\ar @{-} [#1,0]}

\newcommand{\gate}[1]{*{\xy *+<.6em>{#1};p\save+LU;+RU **\dir{-}\restore\save+RU;+RD **\dir{-}\restore\save+RD;+LD **\dir{-}\restore\POS+LD;+LU **\dir{-}\endxy} \qw}

\newcommand{\measureD}[1]{*{\xy*+=+<.5em>{\vphantom{\rule{0em}{.1em}#1}}*\cir{r_l};p\save*!R{#1} \restore\save+UC;+UC-<.5em,0em>*!R{\hphantom{#1}}+L **\dir{-} \restore\save+DC;+DC-<.5em,0em>*!R{\hphantom{#1}}+L **\dir{-} \restore\POS+UC-<.5em,0em>*!R{\hphantom{#1}}+L;+DC-<.5em,0em>*!R{\hphantom{#1}}+L **\dir{-} \endxy} \qw}

\newcommand{\multimeasureD}[2]{*+<1em,.9em>{\hphantom{#2}}\save[0,0].[#1,0];p\save !C *{#2},p+LU+<0em,0em>;+RU+<-.8em,0em> **\dir{-}\restore\save +LD;+LU **\dir{-}\restore\save +LD;+RD-<.8em,0em> **\dir{-} \restore\save +RD+<0em,.8em>;+RU-<0em,.8em> **\dir{-} \restore \POS !UR*!UR{\cir<.9em>{r_d}};!DR*!DR{\cir<.9em>{d_l}}\restore \qw}
\newcommand{\control}{*!<0em,.025em>-=-{\bullet}}
\newcommand{\controlo}{*-<.21em,.21em>{\xy *=<.59em>!<0em,-.02em>[o][F]{}\POS!C\endxy}}
\newcommand{\ctrl}[1]{\control \qwx[#1] \qw}
\newcommand{\ctrlo}[1]{\controlo \qwx[#1] \qw}

\newcommand{\multigate}[2]{*+<1em,.9em>{\hphantom{#2}} \qw \POS[0,0].[#1,0];p !C *{#2},p \save+LU;+RU **\dir{-}\restore\save+RU;+RD **\dir{-}\restore\save+RD;+LD **\dir{-}\restore\save+LD;+LU **\dir{-}\restore}
\newcommand{\ghost}[1]{*+<1em,.9em>{\hphantom{#1}} \qw}

\newcommand{\gategroup}[6]{\POS"#1,#2"."#3,#2"."#1,#4"."#3,#4"!C*+<#5>\frm{#6}}

\newcommand{\lstick}[1]{*!R!<.5em,0em>=<0em>{#1}}

\newcommand{\Qcircuit}[1][0em]{\xymatrix @*[o] @*=<#1>}

\newcommand{\pureghost}[1]{*+<1em,.9em>{\hphantom{#1}}}
\newcommand{\multiprepareC}[2]{*+<1em,.9em>{\hphantom{#2}}\save[0,0].[#1,0];p\save !C
  *{#2},p+RU+<0em,0em>;+LU+<+.8em,0em> **\dir{-}\restore\save +RD;+RU **\dir{-}\restore\save
  +RD;+LD+<.8em,0em> **\dir{-} \restore\save +LD+<0em,.8em>;+LU-<0em,.8em> **\dir{-} \restore \POS
  !UL*!UL{\cir<.9em>{u_r}};!DL*!DL{\cir<.9em>{l_u}}\restore}
\newcommand{\prepareC}[1]{*{\xy*+=+<.5em>{\vphantom{#1\rule{0em}{.1em}}}*\cir{l^r};p\save*!L{#1} \restore\save+UC;+UC+<.5em,0em>*!L{\hphantom{#1}}+R **\dir{-} \restore\save+DC;+DC+<.5em,0em>*!L{\hphantom{#1}}+R **\dir{-} \restore\POS+UC+<.5em,0em>*!L{\hphantom{#1}}+R;+DC+<.5em,0em>*!L{\hphantom{#1}}+R **\dir{-} \endxy}}
\newcommand{\poloFantasmaCn}[1]{{{}^{#1}_{\phantom{#1}}}}
 \layoutstyle{6x9} \indentcaption{1.2em}
\captionnamefont{\bf}\captiondelim{\bf.~}
\captiontitlefont{\rmfamily}\precaption{\hskip 1em\footnotesize {\bf FIGURE~}}
\def\eg{{e.~g.} }\def\ie{{i.~e.} }\def\paragraph#1{\medskip\par\noindent{\bf #1}}
\def\transp#1{{#1}^{T}}\def\map#1{{\mathcal #1}}
\def\K#1{\left|#1\right)\!}\def\B#1{\left(#1\right|}\def\SC#1#2{\left(#1\left|\right.\!#2\right)\!}
\def\trnsfrm#1{\mathscr #1}\def\lntrnsfrm#1{\mathrm #1} \def\tA{\trnsfrm A}\def\tB{\trnsfrm
  B}\def\tC{\trnsfrm C}\def\tD{\trnsfrm D}\def\tE{\trnsfrm E}\def\tF{\trnsfrm F}\def\tG{\trnsfrm G}
\def\tI{\trnsfrm I}\def\tS{\trnsfrm S}
\def\rA{\lntrnsfrm A}\def\rB{\lntrnsfrm B}\def\rC{\lntrnsfrm C} \def\rD{\lntrnsfrm D}\def\rE{\lntrnsfrm
  E}\def\rF{{\lntrnsfrm F}} \def\rG{{\lntrnsfrm G}}\def\rI{\lntrnsfrm I}\def\rH{\lntrnsfrm H}
\def\rL{{\lntrnsfrm L}}\def\rM{\lntrnsfrm M}\def\rN{\lntrnsfrm N}\def\rO{{\lntrnsfrm O}}
\def\rP{\lntrnsfrm P} \def\rX{{\lntrnsfrm X}}
\def\<{\langle}\def\>{\rangle}\def\Reals{\mathbb R}\def\Cmplx{\mathbb C}
\def\Stset{{\mathfrak S}}\def\Trnset{{\mathfrak T}}\def\Cntset{{\mathfrak E}} 
\def\set#1{{\sf #1}}\def\Span{\set{Span}}\def\Erays{\set{Erays}}
\def\n#1{|\!|#1|\!|}\renewcommand{\leq}{\leqslant}
\def\d{\operatorname{d}}
\def\Postulate#1#2#3{\medskip\par\noindent{\bf Postulate #1: #2.$\;$}{\em #3}\medskip\par}

\begin{document}
\title[... a computational grand-unification]{On the ``principle of the quantumness'',\\ the
  quantumness of Relativity,\\ and the computational grand-unification\footnote{Work presented at
    the conference {\em Quantum Theory: Reconsideration of Foundations, 5} held on 14-18 June 2009
    at the International Center for Mathematical Modeling in Physics, Engineering and Cognitive
    Sciences, V\"axj\"o University, Sweden. The ideas on QCFT have been added after the conference.}
}  \author{Giacomo Mauro D'Ariano}{address={{\em QUIT} Group,
    Dipartimento di
    Fisica ``A. Volta'', 27100 Pavia, Italy, {\em http://www.qubit.it}\\
    Center for Photonic Communication and Computing, Northwestern University, Evanston, IL 60208}}
\begin{abstract} I will argue that the proposal of establishing operational foundations of Quantum
  Theory should have top-priority, and that the Lucien Hardy's program on Quantum Gravity should be
  paralleled by an analogous program on Quantum Field Theory (QFT), which needs to be reformulated,
  notwithstanding its experimental success. In this paper, after reviewing recently suggested
  operational ``principles of the quantumness'', I address the problem on whether Quantum Theory and
  Special Relativity are unrelated theories, or instead, if the one implies the other. I show how
  Special Relativity can be indeed derived from causality of Quantum Theory, within the
  computational paradigm ``the universe is a huge quantum computer'', reformulating QFT as a
  Quantum-Computational Field Theory (QCFT).  In QCFT Special Relativity emerges from the fabric of
  the computational network, which also naturally embeds gauge invariance. In this scheme even the
  quantization rule and the Planck constant can in principle be derived as emergent from the
  underlying causal tapestry of space-time.  In this way Quantum Theory remains the only theory
  operating the huge computer of the universe.

  Is the computational paradigm only a speculative tautology (theory as simulation of reality), or
  does it have a scientific value?  The answer will come from Occam's razor, depending on the
  mathematical simplicity of QCFT.  Here I will just start scratching the surface of QCFT, analyzing
  simple field theories, including Dirac's. The number of problems and unmotivated recipes that
  plague QFT strongly motivates us to undertake the QCFT project, since QCFT makes all such problems
  manifest, and forces a re-foundation of QFT.
\end{abstract}
\maketitle
\section{Introduction}
Quantum Theory (QT) is still lacking a foundation. The Lorentz transformations were in the same
situation before the advent of Special Relativity (SR). If one considers the theoretical power of SR
in the ensuing research, one would definitely put the search for an analogous principle of the
``quantumness'' at the top of priorities.  Where such new deeper understanding of QT could lead us?
To a theory of Quantum Gravity---Lucien Hardy would say. Or, even to a more profound understanding
of the whole Physics, since, as I will argue, in a sense QT is the whole Physics.  Should we risk
our research time and money in these hazardous investigations? My answer is a definite: Yes, we
should! Besides, if we take an operational approach, we will stay far away from speculations \eg on
the number of curled-up dimensions of the world or on whether particles are indeed strings or
membranes.  And, in any case, we will end up with a deeper understanding of the relations between
general issues as local observability, no-signaling, locality, causality, local causality,
experimental complexity, computational power, reversibility, and more.

In the last six years I spent a great deal of my time seeking a principle of the ``quantumness'',
and found indeed more than one set of combined principles. Yet, I have not exactly QT in my hands,
but something slightly more general. In the recent article \cite{myCUP2009} I proposed some
postulates that are the basic requirements for operational control and reduction of experimental
complexity, such as causality, local observability, and the existence of states that allow local
calibrations of instruments and local preparation of joint states. These postulates have shown an
unexpected power, excluding all known probabilistic theories except QT. More recently with G.
Chiribella and P.  Perinotti \cite{Chiribella2009unp} we discovered the full potential of a
purifiability postulate, which narrows the probabilistic theory to something very close to QT. We
started with a new fresh approach that turned out to be very efficient, as it provides a
``diagrammatic'' way of proving theorems. From the postulates we derived most of the relevant
features of QT and Quantum Information, including dilation theorems, error correction teleportation,
no-cloning, no-bit-commitment, etc. In Sect. \ref{s:opframe} I will briefly review this axiomatic
excursus.

Writing a conference proceedings for an invited talk gives me the irresistible opportunity of adding
something more of what I said at the conference. I therefore decided to spend the second part of the
paper for taking my first move from Quantum Theory toward Quantum Field Theory. I've been always
interested in questions as: Where Special Relativity comes from? Are Quantum Theory and Special
Relativity unrelated theories? Is Quantum Field Theory an additional theoretical layer over QT and
SR?  Where the quantization rules and the Planck constant come from? Here I will argue that a
possible answer to all these questions is provided by the ``computational paradigm'': {\em the
  universe is a huge quantum computer}.\footnote{This can be also regarded as an ultra-strong
  version of the Turing test: {\em is reality indistinguishable from a perfect quantum simulation of
    it?}} I will take the paradigm seriously as a theoretical framework, and analyze the
implications and problems posed. I will show how, amazingly, from the fabric of the computational
network, space and time emerge naturally endowed with the relativistic covariance, just as a
consequence of local causality (\ie the gates involve only a finite number of systems) and of
uniformity and isotropy of the computational circuit.  What a field will look in this description? A
classical field will be a classical computation of an input string of bits (or $d$its), a quantum
field a quantum computation of a string of qubits (or qudits).  Different fields result from
different choices of the circuit gates. The gauge-invariance will be simply an arbitrary choice of
basis at the gates inputs. The quantization rule itself---which defines the classical
Schr\H{o}dinger field---can in principle be written inside the gates themselves. We will see that
the discrete computational framework makes the Zitterbewegung of the Dirac particle a general
phenomenon within QCFT, the zig-zag frequency being related to the particle mass, whereas the Planck
constant resorts being the blurring scale of the causal network, via the Compton wavelength of the
particle.  In the paper I will work out some preliminary exercises in ``Quantum-Computational Field
Theory'' (QCFT), providing the circuit implementation of simple field theories, e.g. the
Klein-Gordon and the Dirac fields in on space-dimension.

Is QCFT only a speculative tautology, or does it have a scientific value?  Before answering we first
need to see what of QFT is possible to derive as coarse-graining of QCFT. Apart from a matter of
taste related to the computational circuit as an ontology, the two crucial criteria will be Occam
razor and mathematical simplicity. I must however emphasize that in any case the QCFT program is a
must for the following reasons.  First, QCFT solves a number of logical and mathematical problems
that plague QFT \cite{Cao1997,Teller1995}, besides allowing a unified framework for different
fields, giving a mechanism for relativistic invariance, and, last but not least, providing a
systematic way for consistently generalizing the whole theoretical framework in view of Quantum
Gravity, with the possibility of changing the computational engine from QT to a super-quantum
operational theory, or even an input-output network with no pre-established causal relations.  All
these nice features may motivate adopting QCFT in place of QFT, QFT being still not well founded
both operationally and logically (see \eg quantization rule, Feynman path integral, Grassman
variables, microcausality...)  We will discuss more about these issues at the end of the paper.
Another reason for exploring QCFT is that QCFT represents the first test of the Lucien Hardy's
program of an operational approach to Quantum Gravity. In fact, before building up a theory of
Quantum Gravity, we first should check the approach against a well assessed phenomenology, such as
that of particle physics: this would also be much easier than deriving a theory of Quantum Gravity.
QCFT would also bring the powerful point of view of Quantum Information inside the world of particle
physics.

\section{The operational framework.\label{s:opframe}}
The starting point of the operational framework is the notion of {\bf test}. A test is made of the
following ingredients: a) a complete collection of {\bf outcomes}, b) input {\bf systems}, c) output
systems. It is represented in form of a box, as follows

$$\Qcircuit @C=1em @R=.7em @! R {
&\qw\poloFantasmaCn{\rA_1}&\multigate{1}{\{\tA_i\}}&\qw\poloFantasmaCn{\rB_1}&\qw\\
&\qw\poloFantasmaCn{\rA_2}&\ghost{\{\tA_i\}}&\qw\poloFantasmaCn{\rB_2}&\qw}\qquad
\Qcircuit @C=1em @R=.7em @! R {
&\qw\poloFantasmaCn{\rA_1}&\multigate{1}{\tA}&\qw\poloFantasmaCn{\rB_1}&\qw\\
&\qw\poloFantasmaCn{\rA_2}&\ghost{\tA}&\qw\poloFantasmaCn{\rB_2}&\qw}$$

The left wires represent the input systems, the right wires the output systems, and $\{\tA_i\}$ the
collection of outcomes.  We often represent not the complete test, but just a single outcome
$\tA_i$, or, more generally, a subset $\tA\subset\{\tA_i\}$ of the collection of outcomes, \ie an
{\bf event}, as in the right box in figure.  The number of wires at the input and at the output can
vary, and one can have also no wire at the input and/or at the input. Depending on the context, the
test can be regarded as a man-made apparatus or as a nature-made physical interaction. The set of
events of a test is closed under union (also called {\bf coarse-graining}), intersection, and
complementation, thus making a Boolean algebra. A {\bf refinement} of an event $\tA$ is a set of
events $\{\tA_i\}$ occurring in some test such that $\tA=\cup_i \tA_i$.  Generally an event can have
different refinements depending on the test to which it belongs, or it may be unrefinable within
some test.  An event that is unrefinable within any test is called {\bf atomic}.

The natural place for a test/event is inside a network of other tests/events, and to understand the
origin of the box representation and the intimate meaning of the test/event you should regard it
connected to other tests/events in a circuit, \eg as follows

{\footnotesize $$
\Qcircuit @C=1em @R=.7em @! R {
  \multiprepareC{3}{\Psi}&\qw\poloFantasmaCn{\rA}&\multigate{1}{\tA}&\qw\poloFantasmaCn{\rB}&\gate{\tC}&\qw\poloFantasmaCn{\rC}&\multigate{1}{\tE}&\qw\poloFantasmaCn{\rD}&\multimeasureD{2}{\tG}\\
  \pureghost{\Psi}&\qw\poloFantasmaCn{\rE}&\ghost{\tA}&\qw\poloFantasmaCn{\rF}&\multigate{1}{\tD}&\qw\poloFantasmaCn{\rG}&\ghost{\tE}&&\pureghost{\tG}\\
  \pureghost{\Psi}&\qw\poloFantasmaCn{\rH}&\multigate{1}{\tB}&\qw\poloFantasmaCn{\rL}&\ghost{\tD}&\qw\poloFantasmaCn{\rM}&\multigate{1}{\tF}&\qw\poloFantasmaCn{\rN}&\ghost{\tG}\\
  \pureghost{\Psi}&\qw\poloFantasmaCn{\rO}&\ghost{\tB}&\qw\poloFantasmaCn{\rP}&\qw&\qw &\ghost{\tF}\\
}$$} 

The different letters $\rA,\rB,\rC,\ldots$ labeling the wires denote different ``types of system''.
We can connect only an input wire of a box with an output wire of another box, the two wires
having the same label. Loops are forbidden. Among the different kinds of systems, we have a special
one called {\bf trivial system}, denoted by $\rI$, which we conveniently represent by no wire, but
instead, by drawing the corresponding side of the box convexly rounded as follows $\Qcircuit @C=1em
@R=.7em @! R {\prepareC{\omega}&\qw\poloFantasmaCn{\rA}&\qw}:=\Qcircuit @C=1em @R=.7em @! R { &\qw
  \poloFantasmaCn{\rI}&\gate{\omega}&\qw\poloFantasmaCn{\rA}&\qw}$, and $\Qcircuit @C=1em @R=.7em @!
R{&\qw\poloFantasmaCn{\rA}&\measureD{a}}:=\Qcircuit @C=1em @R=.7em @!  R
{&\qw\poloFantasmaCn{\rA}&\gate{a}&\qw\poloFantasmaCn{\rI}&\qw}$.

The fact that there are no closed loops gives to the circuit the structure of a DAG (directed
acyclic graph), with vertices corresponding to tests/events, and edges to wires.  The absence of
closed loops corresponds to the requirement that the test/event is one-use only. We also must keep
in mind that there are no constraints for disconnected parts of the network, which can be arranged
freely (this would not be true \eg for a quaternionic quantum network). Finally, we will also
consider \emph{conditioned tests}, where one can choose a different test depending on the outcome of
a test connected to the input. The construction of the network mathematically is equivalent to the
construction of a \emph{symmetric strict monoidal category} (see Ref.\cite{Abramsky2004p3991}).

\medskip In order to make predictions about the occurrence probability of events based on current
knowledge, one needs a ``theory''. {\em An {\bf operational theory}} \cite{Chiribella2009unp} {\em is
  specified by a collection of systems, closed under parallel composition, and by a collection of
  tests, closed under parallel/sequential composition and under randomization. The operational
  theory is \emph{probabilistic} if every test from the trivial system to the trivial system is
  associated to a probability distribution of outcomes.}

Therefore a probabilistic theory provides us with the joint probabilities for all possible events
for any closed network (namely with no overall input and output). The probability itself will be
conveniently represented by the corresponding network of events. We must keep in mind that the
probability of an event is independent on the test to which it belongs, and this legitimates using
networks of events without specifying the test. In the following, we will denote the
set of events from system $\rA$ to system $\rB$ as $\Trnset(\rA,\rB)$, and use the abbreviation
$\Trnset(\rA):=\Trnset(\rA,\rA)$.

Two wires in a circuit are {\bf input-output adjacent} if they are the input and the output of the
same box. By following input-output adjacent wires in a circuit in the input-to-output direction we
draw an {\bf input-output chain}. Two systems (wires) that are not in the same input-output chain
are called {\bf independent}. A set of pairwise independent systems is a {\bf slice}. The slice is
called global if it partitions the circuit into two parts.

By construction it is clear that a global slice always partitions a closed bounded circuit into two
parts: a preparation test and an observation test.  Thus, a diagram of the form $\Qcircuit @C=1em
@R=.7em @! R { \prepareC{\tA_i}&\qw&\qw\poloFantasmaCn{\rA}&\qw &\measureD{\tB_j}}$ generally
represents the event corresponding to an instance of a concluded experiment, which starts with a
preparation and ends with an observation. The probability of such event will be denoted as
$\SC{\tB_j}{\tA_i}$, using the ``Dirac-like'' notation, with {\em rounded ket} $\K{\tA_i}$ and {\em
  bra} $\B{\tB_j}$ for the preparation and the observation tests, respectively.  In the following we
will use lowercase  Greek letters for preparation events, and lowercase Latin letters for observation
events. The following notations are equivalent: $\B{a}\tA\K{\rho}= \Qcircuit @C=1em @R=.7em @! R {
  \prepareC{\rho} &\gate{\tA}&\measureD{a}}$, $\Qcircuit @C=1em @R=.7em @! R
{&\gate{\tA}&\measureD{a}}= \Qcircuit @C=1em @R=.7em @! R {&\measureD{a\circ\tA}}$, and
$\B{a}\tA=\B{a\circ\tA}$. The event $\tA$ can be regarded as ``transforming'' the observation event
$a$ into the event $a\circ\tA$. The same can be said for the preparation event.  The sets of
preparation and observation tests for system $\rA$ will be denoted as $\Stset(\rA)$ and
$\Cntset(\rA)$, respectively.
 
\subsection{States, effects, transformations}

In a probabilistic theory, a preparation-event $\rho_i$ for system $\rA$ is
naturally identified with a function sending observation-events of $\rA$ to probabilities, namely
\begin{equation}\label{statfun}
\rho_i: \Stset(\rA)\to [0,1], \quad \B{a_j}\mapsto \SC{a_j}{\rho_i}.
\end{equation}
Similarly, observation-events are identified with functions from preparation-events to
probabilities
\begin{equation}
a_j: \Cntset(\rA)\to [0,1], \quad \K{\rho_i}\mapsto \SC{a_j}{\rho_i}.  
\end{equation}    
Considered as probability rule, two observation-events (preparation-events) corresponding to the
same function are indistinguishable. We will then call {\bf states} the equivalence classes of
indistinguishable preparation-events, and {\bf effects} the equivalence classes of indistinguishable
observation-events, and keep the same notation used for events $\Stset(\rA)$ and $\Cntset(\rA)$ for
the respective equivalence classes. According to our definition states are necessarily separating
for effects, and viceversa effects are separating for states.

Since states (effects) are functions from effects (states) to probabilities, one can take linear
combinations of them. This defines the real vector spaces $\Stset_\Reals (\rA)$ and $\Cntset_\Reals
(\rA)$, dual each other, and one has $\dim (\Stset_\Reals (\rA)) = \dim (\Cntset_\Reals (\rA))$.
Linear combinations with positive coefficients define the two convex cones $\Stset_{+} (\rA)$ and
$\Cntset_{+} (\rA)$, dual each other. Moreover, since the experimenter is free to randomize the
choice of devices with arbitrary probabilities, all sets $\Stset (\rA)$, $\Cntset (\rA)$ and
$\Trnset(\rA)$ are convex.  Linearity is naturally transferred to any kind of event, via linearity
of probabilities. Moreover, every event $\tA\in\Trnset(\rA,\rB)$ induces a map from $\Stset(\rA\rC)$
to $\Stset(\rB\rC)$ for every system $\rC$, uniquely defined by
\begin{equation}\label{extensA} 
\tA :\K{\rho}_{\rA \rC} \in \Stset (\rA \rC) \mapsto (\tA\otimes \tI_\rC) \K{ \rho}_{\rA \rC} \in
\Stset (\rB \rC),  
\end{equation}  
$\tI_\rC$ denoting the identity transformation on system $\rC$. The map is linear from
$\Stset_\Reals(\rA\rC)$ to $\Stset_\Reals (\rB\rC)$.  Operationally two events $\tA$ and $\tA'$ are
indistinguishable if for every possible system $\rC$ they induce the same map, and we will call {\bf
  transformations} from $\rA$ to $\rB$ the equivalence classes of indistinguishable events from
$\rA$ to $\rB$. Henceforth, we will identify events with transformations, and accordingly, a test
will be a collection of transformations.

In the following, if there is no ambiguity, we will drop the system index to the identity event.
Notice that generally two transformations $\tA,\tA' \in \Trnset (\rA , \rB)$ can be different even
if $\tA\K{ \rho}_\rA = \tA' \K{ \rho}_\rA$ for every $\rho \in \Stset (\rA)$. Indeed one has
$\tA\neq\tA'$ if that there exists an ancillary system $\rC$ and a joint state $\K{\rho}_{\rA \rC}$
such that
$(\tA \otimes \tI)\K{\rho}_{\rA \rC} \not = (\tA'\otimes\tI)\K{ \rho}_{\rA \rC}$.
We will come back on this point when discussing local discriminability.

\subsection{The postulates.}
In the networks discussed until now we had sequences of tests, however, such sequences were not
necessarily {\em temporal}, or \emph{causal}. We have shown that every portion of a closed network
is equivalent to a preparation test connected to an observation test. The causal condition can now
be formulated as follows:

\paragraph{Causal Condition.} \cite{Chiribella2009unp} {\em A theory is \emph{causal} if every
  preparation-event $\K{\rho_j}_\rA$ has a probability $p(\rho_j)$ that is independent on the choice
  of test following the preparation test. Precisely, if $\{\tA_i\}_{i \in \rX}$ is an arbitrary test
  from $\rA$ to $\rB$, one has $p(\rho_j) = \sum_{i\in\rX}p(\tA_i\rho_j)$.}

In Ref. \cite{myCUP2009} the causality condition has been introduced as the asymmetry of
marginalization of the joint probability of two input-output contiguous tests, with the input
marginal independent on the choice of the output test, but not viceversa. It is easy to see that
such condition (also called {\em no-signaling from the future}) is equivalent to the present causal
condition.

Notice that there exist indeed input-output relation that have no causal interpretation. A concrete
example of such theories is that considered in Refs.  \cite{supermaps,comblong}, where the states
are quantum operations, and the transformations are ``supermaps'' transforming quantum operations
into quantum operations. In this case, transforming a state means inserting the quantum operation in
a larger circuit, and the sequence of two transformation is not causal. In Ref.
\cite{upcomingPaolo} the operational non causal framework is thoroughly analyzed: this may
constitute a crucial ingredient for conceiving a quantum theory of gravity, as proposed in Ref.
\cite{Hardy2007p2038}.  The causality principle naturally leads to the notion of conditioned
tests, generalizing both notions of sequential composition and randomization of tests. For a precise
definition see Ref. \cite{Chiribella2009unp}.

Causal theories have a simple characterization as follows \cite{Chiribella2009unp}: {\em A theory
  is causal if and only if for every system $\rA$ there is a unique deterministic effect
  $\B{e}_\rA$}. Equivalently: {\em A theory where every state is proportional to a deterministic
  preparation test is causal.}

When considering a causal theory, we can define the notion of {\bf marginal state} of $\K\sigma_{\rA
  \rB}$ on system $\rA$ as the state $\K \rho_\rA := \B e_\rB \K\sigma_{\rA \rB}$.

In the following, when considering a transformation in $\tA\in\Trnset(\rA,\rB)$ acting on a joint
state $\omega\in\Stset(\rA\rC)$, we will think the transformation as acting on $\omega$ locally, namely
we will use the following natural abbreviations $\tA\omega\equiv(\tA\otimes\tI)\K{\omega}_{\rA\rC}$
and $\omega(\tA)\equiv \B{e}_{\rA\rC}(\tA\otimes\tI)\K{\omega}_{\rA\rC}$ for
$\tA\in\Trnset(\rA,\rB),\omega\in\Stset(\rA\rC)$. For probabilities the abbreviation corresponds
to take the marginal state.

Causality implies the impossibility of signalling without exchanging systems
\cite{Chiribella2009unp}: 

\bigskip{\bf Theorem (No signalling without exchange of physical
    systems)} {\em In a causal theory it is impossible to have signalling
  without exchanging systems.}
\bigskip

The second main assumption on the probabilistic theory is: 

\paragraph{Local discriminability:} {\em A theory satisfies local discriminability if every couple
  of different states $\rho, \sigma \in \Stset (\rA \rB)$ can be discriminated locally, namely if
  there are two local effects $a \in \Cntset (\rA)$ and $b \in \Cntset (\rB)$ such that}
{\footnotesize $$
\begin{matrix}\Qcircuit @C=1em @R=.7em @! R {\multiprepareC{1}{\rho}& \qw \poloFantasmaCn \rA
    &\measureD a \\  \pureghost\rho & \qw \poloFantasmaCn \rB &\measureD b}\end{matrix}
\;\begin{matrix}\not =\end{matrix}\;
\begin{matrix}\Qcircuit @C=1em @R=.7em @! R {\multiprepareC{1}{\sigma}& \qw
      \poloFantasmaCn \rA &\measureD a \\  \pureghost\sigma & \qw \poloFantasmaCn \rB &\measureD b}
  \end{matrix}$$} Another way of stating local discriminability is to say that the set of factorized
effects is separating for the joint states. Local discriminability represents a dramatic
experimental advantage, since, otherwise, one would need to built up a $N$-system test in order to
discriminate an $N$-partite joint state. Instead, thanks to local discriminability, we can recover
the full joint state from just local observations. Local discriminability is equivalent to local
observability, namely the possibility of performing a complete tomography of a multipartite state
using only local tests. A mathematical restatement of local discriminability is the tensor product
rule for states $\Stset_\Reals(\rA\rB)=\Stset_\Reals(\rA)\otimes \Stset_\Reals(\rB)$ and effects 
$\Cntset_\Reals(\rA\rB)=\Cntset_\Reals(\rA)\otimes \Cntset_\Reals(\rB)$.  Another consequence is
that transformations in $\Trnset(\rA,\rB)$ are completely specified by their action only on local
states $\Stset(\rA)$, without the need of considering ancillary extension.

For a causal theory with local discriminability one has a nice Bloch representation of states and
effects, and a matrix representation for transformations as linear maps over them
\cite{dariano-losini2005}. For more details on such representation see Ref. \cite{tosini}.

\medskip\par A state $\Phi\in\Stset(\rA\rB)$ of a bipartite system $\rA\rB$ induces the
cone-homomorphism\footnote{A cone-homomorphism between the cones $K_1$ and $K_2$ is simply a linear
  map between $\Span_\Reals(K_1)$ and $\Span_\Reals(K_2)$ which sends elements of $K_1$ to elements
  of $K_2$, but not necessarily vice-versa.}
$\Trnset_+(\rA)\ni\tA\mapsto(\tA\otimes\tI)\Phi\in\Stset_+(\rA\rB)$. If this is a cone-monomorphism
$\Phi$ is called {\bf dinamically faithful} with respect to $\rA$, since the output state
$(\tA\otimes\tI)\Phi$ is in one-to-one correspondence with the local transformation $\tA$. When it
is a cone-epimorphism the state is called {\bf preparationally faithful} with respect to $\rA$,
since every bipartite state $\Psi$ can be achieved as $\Psi=(\tA_\Psi\otimes\tI)\Phi$ for some local
transformation $\tA_\Psi$.

We can now state the postulate:

\Postulate{PFAITH}{Existence of a symmetric preparationally faithful pure state}{For any couple of
  identical systems, there exists a symmetric (invariant under permutation of the two systems)
  bipartite state which is both pure and preparationally faithful.}

Postulate PFAITH is central in the operational probabilistic theories of Refs.
\cite{myCUP2009,Chiribella2009unp}, since it concerns the possibility of calibrating any test and
of preparing any joint bipartite state only by means of local transformations. The postulate leads
to many relevant features of the theory \cite{myCUP2009}. For a bipartite system $\rA\rB$ with
identical systems $\rA=\rB$ (in the following we simply will write $\rA\rA$ instead of $\rA\rB$),
upon denoting by $\Phi\in\Stset(\rA\rA)$ the symmetric preparationally faithful state, the postulate
implies that: (1) $\Phi$ is also dinamically faithful with respect to both systems, whence the state
achieves the cone-isomorphism\footnote{Two cones $K_1$ and $K_2$ are isomorphic iff there exists a
  linear bijective map between the linear spans $\Span_\Reals(K_1)$ and $\Span_\Reals(K_2)$ that is
  cone preserving in both directions, namely it and its inverse map must send $\Erays(K_1)$ to
  $\Erays(K_2)$ and positive linear combinations to positive linear combinations.}
$\tA\in\Trnset_+(\rA)\mapsto(\tA\otimes\tI)\Phi\in\Stset_+(\rA\rA)$; (2) The state
$\Psi=(\tA_\Psi\otimes\tI)\Phi$ is pure iff $\tA_\Psi$ is atomic; (3) the theory is weakly
self-dual, namely one has the cone-isomorphism $\Cntset_+(\rA)\simeq\Stset_+(\rA)$ induced by the
map $\Phi(a,\cdot)=\omega_a$ $\forall a\in\Cntset_+(\rA)$; being $\Phi$ a cone-isomorphism for both
$\rA$ and $\rB$, we can operationally define the {\bf transposed transformation}
$\tA^\prime\in\Trnset_\Reals(\rA)$ of $\tA\in\Trnset_\Reals(\rA)$ through the identity
$(\tA^\prime\otimes\tI)\Phi=(\tI\otimes\tA)\Phi$; (4) the identical transformation $\tI$ is atomic;
(5) the transposed of a physical automorphism of the set of states is still a physical automorphism
of the set of states; (6) the maximally chaotic state $\chi:=\Phi(e,\cdot)$ is invariant under the
transpose of a channel (deterministic transformation) whence, in particular, under a physical
automorphism of the set of states.

A stronger version of PFAITH, satisfied by Quantum Theory, requires the existence of a symmetric
preparationally {\bf superfaithful} state $\Phi$ such that also $\Phi\otimes\Phi$ is preparationally
faithful, whence $\Phi^{\otimes 2n}$ is preparationally faithful with respect to $\rA^n$, $\forall
n>1$.

\medskip\par The additional Postulate FAITHE \cite{myCUP2009} makes the probabilistic theory closer
to Quantum Theory. Since a preparationally faithful state is also dynamically faithful, it is
indeed an isomorphism, and, as a matrix, it is invertible. However, generally its inverse is not a
bipartite effect. This is exactly what postulate FAITHE requires, namely
\Postulate{FAITHE}{Existence of a faithful effect}{There exists a bipartite effect $F$ achieving
  probabilistically the inverse of the cone-isomorphism $\Cntset_+(\rA)\simeq\Stset_+(\rA)$ given by
  $a\leftrightarrow\omega_a:=\Phi(a,\cdot)$, namely
  $\B{F}_{23}\K{\omega_a}_2=\B{F}_{23}\B{a}_1\K{\Phi}_{12}=\alpha \B{a}_3$, $0<\alpha\leq 1$.  }
This is equivalent to $\B{F}_{23}\K{\Phi}_{12}=\alpha\tS_{13}$, $\tS_{ij}$ denoting the
transformation which swaps the $i$th system with the $j$th system, namely the {\bf probabilistic
  teleportation}.  One has $\alpha=\B{e}_{14}\B{F}_{23}\K{\Phi}_{12}\K{\Phi}_{34}$, and the maximum
value of $\alpha$ (over all bipartite effects) depends on the particular probabilistic theory. It is
easy to show \cite{tosini} that if a probabilistic theory does not satisfy Postulate FAITHE, then
teleportation is impossible.  Also, one can see\cite{myCUP2009} that the existence of a
superfaithful states implies Postulate FAITHE.

  In Ref. \cite{Chiribella2009unp} the purifiability of the states of the theory has been
  considered as a postulate. More precisely, in its stronger form the postulate is:
  \Postulate{PURIFY}{Purifiability of all states}{For every state $\omega\in\Stset(\rA)$ there
    exists a purification $\Omega\in\Stset(\rA\rA)$, namely a state $\Omega$ having $\omega$ as
    marginal state, \ie $\B{e}_2\K{\Omega}_{12}=\K{\omega}_1$. The purification is unique up to
    reversible channels on the purifying system.}

  Postulate PURIFY has an amazing list of consequences \cite{Chiribella2009unp}, which narrow the
  probabilistic theory to something very close to Quantum Theory. First of all it entails postulates
  PFAITH and FAITHE. Then, it leads to all main theorems of Quantum Theory and Quantum Information,
  including dilation theorems, error correction, no-cloning, teleportation, no-bit-commitment, etc.
  The dilation theorem for channels is equivalent to PURIFY, thus providing its interpretation as
  ``irreversibility as lack of control'', since each channel possesses a reversible dilation.
  Another power of the postulate PURIFY is that it allows to derive all proofs by purely
  diagrammatic identities. It would be too long to review all implications of the postulate, and the
  reader is addressed to the original publication \cite{Chiribella2009unp}.

\section{What's next? A computational 
grand unification} 

Are Quantum Theory and Special Relativity unrelated theories? Is Quantum Field Theory an additional
theoretical layer over them? Where the quantization rules and the Planck constant come from? As I
mentioned in the introduction, a possible answer to all these questions is to consider a field as a
large quantum computation.  Let's then take this new paradigm seriously, and see what we can get out
of it. First, let's see how space-time and Lorentz invariance emerge from the computational circuit.
\subsection{How space-time and Special Relativity emerge from the circuit}
The Lorentz transformations are the most general change of global reference frame obeying the
Galileo relativity principle, which include homogeneity and isotropy of space and homogeneity of
time. The derived transformations depend on a constant parameter with the dimensions of a velocity,
independent on the relative speed of the two reference systems and separating velocities into two
disconnected regions. From our experience such constant is the speed of light, which bounds
velocities from above.\footnote{The first Einstein's Postulate, namely the Galileo's principle (the
  physical laws are the same in all inertial reference frames) implies isotropy and homogeneity of
  space and homogeneity of time, whence it is sufficient to derive the Lorentz transformations.
  Thus the Lorentz transformations can be derived solely from the Einstein's first Postulate,
  without using the second one (the speed of light is the same in every inertial frame of
  reference). Therefore, the Lorentz transformations are just a consequence of the Galileo principle. This
  fact has been known for long time (see Ref.  \cite{Lee1975p3778}). }  I will now show how
space-time and the Lorentz transformations arise in a quantum computational circuit.

We should regard space and time not as pre-conceptions, but as an efficient model of causal
relations between different events that we (can potentially) experiment. Operationally space and
time make no sense without events ``inside'' them, and in order to measure them we need special sets
of events---``meters'' and ``clocks''---that we assume as universal. Let's now consider a network of
events, connected by causal''wires'', also called ``systems''. If we take ``causality'' as primary
and construct space and time from it, we will recognize as a ``time-direction'' any causal chain of
systems, whereas a ``space-like surface'' will be any global slice made of independent systems. A
set of global slices covering all wires of the circuit will be a ``foliation'', as in the
Tomonaga-Schwinger relativistic approach to Schr\H{o}dinger equation. Clearly both the time and
space directions are arbitrary, reflecting the observer's subjective choice. The specific foliation
along with a set of causal chains covering all wires will establish a reference system.  This will
be the equivalent of a (nonuniform, locally accelerated) reference system. Notice that we have no
metric, since events can be moved and causal wires can be stretched at will. However, we have a
notion of topology, and we can define closeness of two wires, either in time or space direction,
in terms of the number of events through which the systems are causally connected each other, or
through which the two independent systems are causally connected to the same event. 

Even though in the causal network we have no notion of metric, we can count events, and the more
contiguous events in a causal chain the longer the time interval will be, the more contiguous events
in a slice the larger the space interval. We should keep in mind that on each gate we can locate
only one event, whence an infinite global gate corresponds to a space dimension collapsing to a
single point: for this reason gates must involve a number of systems that is finite (or infinite of
a lower order than a global slice).

\begin{figure}[t]
\begin{minipage}{2.5in}
\def\gateshiftx{.9in}\def\gateshifty{-.1in}
\vskip\gateshifty\hskip\gateshiftx
\includegraphics[width=.5\textwidth]{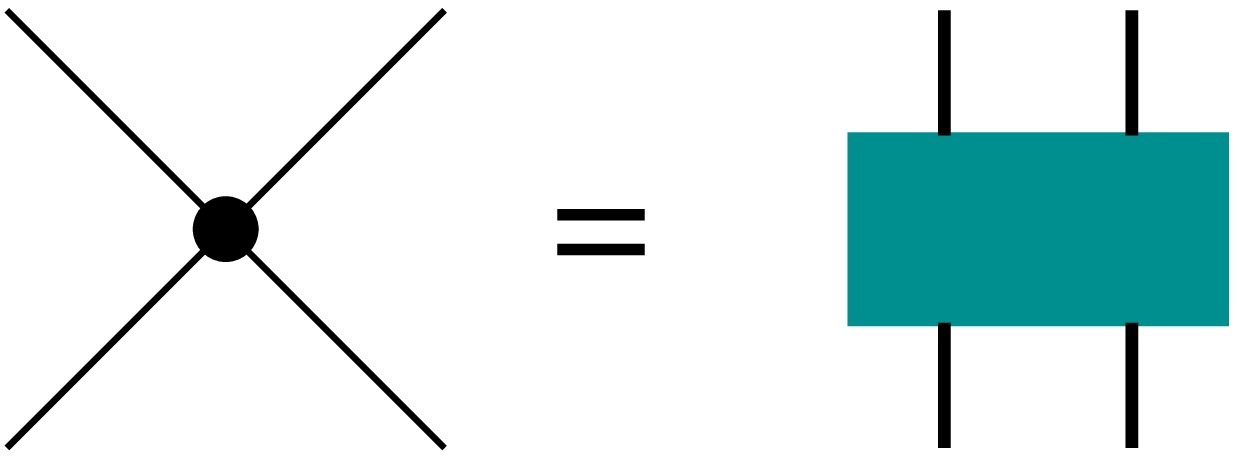}
\hskip -\gateshiftx\vskip -\gateshifty\par\noindent
\includegraphics[width=\textwidth]{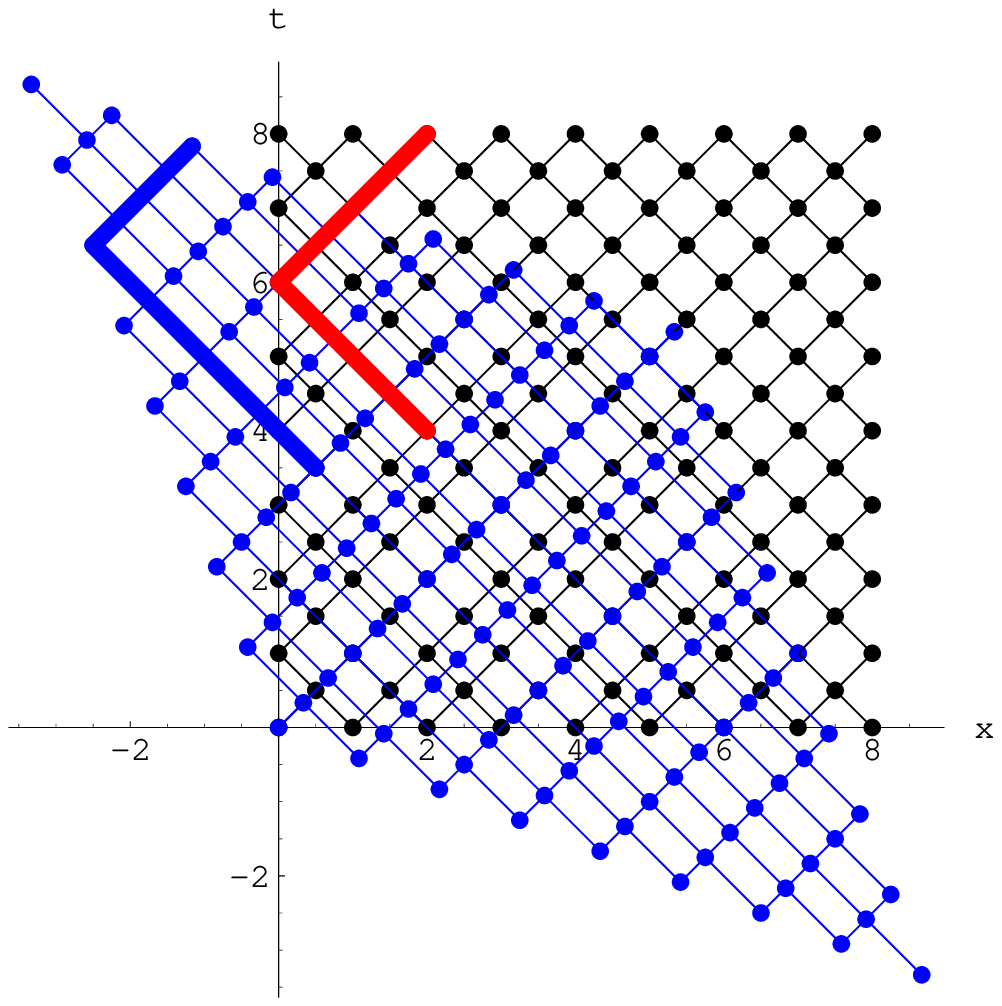}
\end{minipage}
\begin{minipage}{3.3in}
\includegraphics[width=\textwidth]{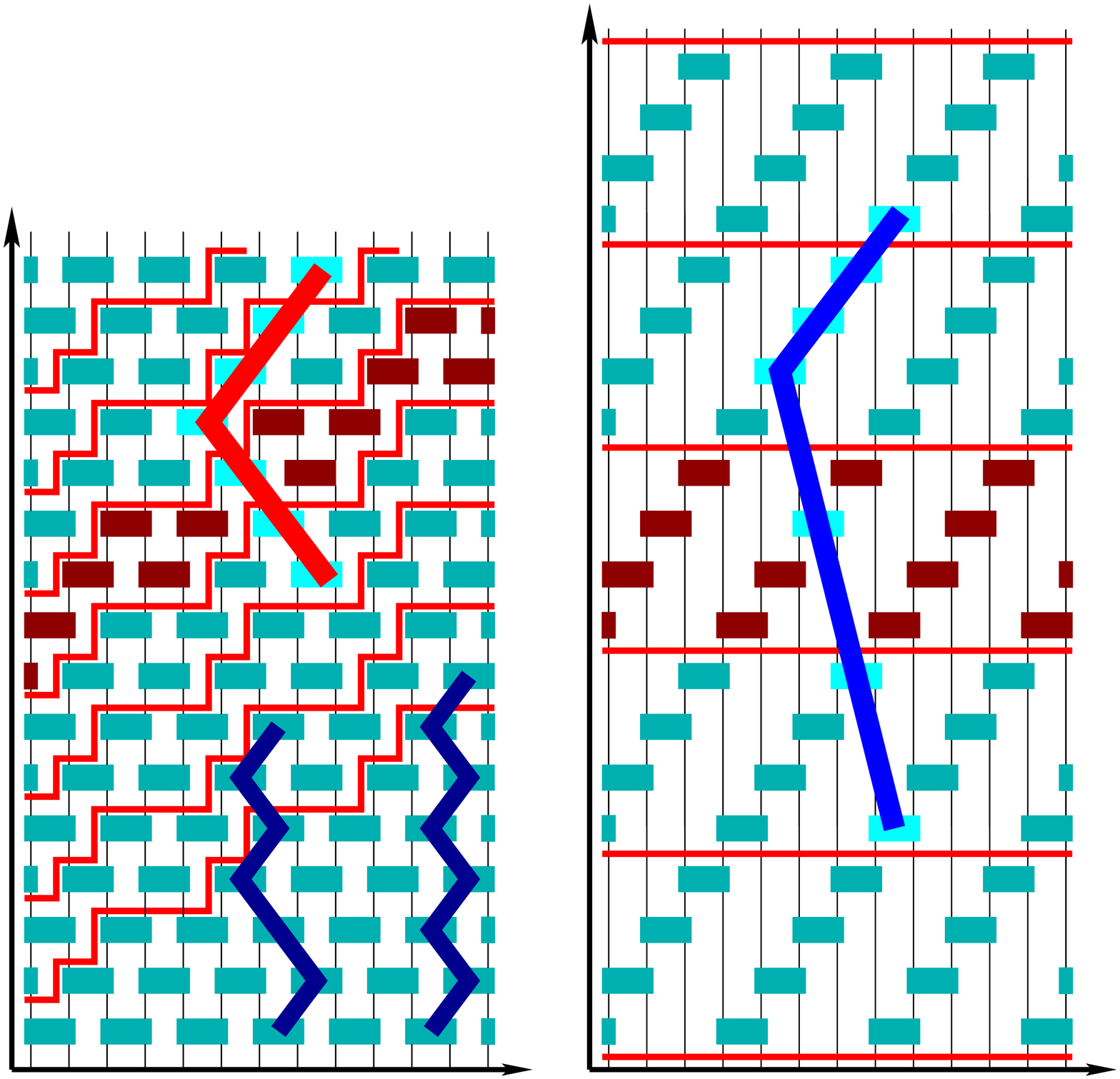}
\end{minipage}
\caption{\rm Illustration on how Special Relativity emerges from the causality of a computational
  network. The networks in the left figure are a convenient DAG representation of the circuits in
  the right figures (see the gate-node correspondence in the top-left inset). {\bf Left figure:} the
  two DAGs are Lorentz-transformed each other ($v=\tfrac{5}{13}c$). The two thick broken lines
  represent a {\em clock tic-tac} made with a light-pulse reflecting between two mirrors. {\bf Right
    figures:} the computational circuit equivalent to the DAG on the left. A clock tic-tac is
  superimposed, along with a global uniform foliation (thin staircase lines). The second circuit is
  obtained from the first one upon stretching wires in such way to put the slices of the foliation
  parallel, to reflect the change of reference system. The corresponding clock tic-tac is reported.
  Notice the similarity with the left figure (apart from a rotation). The Lorentz contraction of
  space emerges as a consequence of the reduced density of events, whereas the time-dilation is
  evident by counting the number of events during the complete tic-tac (the asymmetry between the
  ``tic'' and the ``tac'' is due to the relative motion between the reference frames). The two
  snake-like lines in the center figure depict the {\em Zitterbewegung} phenomenon.\label{fig:synchronous}}
\end{figure}
We are now ready to understand how Lorentz transformations arise in a quantum computational circuit.
Consider a uniform foliation with ``parallel'' slices, as in Fig. \ref{fig:synchronous} (this is the
first observer's modeling of causal relations).  Then, stretch the wires (corresponding to the
identity transformation) to make each slide straight.  In Fig.  \ref{fig:synchronous} it is apparent
how this actually corresponds to a Lorentz transformation, with the Lorentz space-contraction
emerging as a consequence of a resulting reduced density of independent events, whereas
time-dilation comes from a larger number of events during the same tic-tac of a clock. On the other
hand, as already said, the Lorentz transformations can be derived from the Galileo principle only.
We then just need the computational circuit to be both isotropic and homogeneous in space in the
continuum limit, and likewise homogeneous in time, since the bound for speeds simply arises from
finiteness of gates (see how the ``fastest'' causal path has a maximum inclination, beyond which one
needs to go backward in time). As illustrated in Fig. \ref{fig:synchronous}, the {\em maximal causal
  speed} $\upsilon_{caus}$---i.~e. the maximal speed that doesn't violate causality---corresponds to
all gates achieving a complete swapping between systems, and is given by $\upsilon_{caus}= a/\tau$,
where $a$ and $\tau$ are the minimal amount of space and time, respectively, that we attribute to
in-principle discriminable of events, corresponding to the space and time separation of the gates in
the circuit.

\subsection{Simulating quantum fields by computational circuits}
Now that we have seen how space-time and Lorentz transformations emerge from a quantum computational
circuit, let's briefly explore which kind of problems QCFT can pose.  Even though the complete
theory is the second-quantized field theory (QFT${}_2$), it would be interesting to consider also
first-quantized theories (QFT${}_1$). Likewise, for simulating the field theory we can consider a
Second-Quantized Computational Field Theory (QCFT${}_2$) and a First-Quantized one (QCFT${}_1$).  We
know the meaning of QFT${}_1$ and QFT${}_2$.  By QCFT${}_2$ I simply mean the customary
quantum-circuit made of qubits, qudits, or harmonic oscillators. The meaning of QCFT${}_1$ will be
explained in the remaining part of this section. As regards simulating the field theory by a quantum
computational circuit, we have three possibilities: 1) simulating QFT${}_2$ by QCFT${}_2$;
simulating QFT${}_1$ by QCFT${}_2$; simulating QFT${}_1$ by QCFT${}_1$. We cannot reasonably
simulate a QFT${}_2$ by QCFT${}_1$.  Let's now briefly analyze the three possibilities.

\paragraph{Simulating QFT${}_2$ by QCFT${}_2$: the Klein-Gordon equation.}
In a QFT${}_2$ the field is an operator function of space and time. We want now to simulate the
evolution of the field by a quantum computational circuit with gates evolving the field locally. The
field $\phi(x)$ will be described by a set of operators $\phi_n:=a^{\frac{1}{2}}\phi(n a)$, $a$
denoting the infinitesimal space granularity (for simplicity we consider one space-dimension). In
order to mimic the usual quantum field theory, the field operators must satisfy equal-time
commutation/anticommutation relations for the Fermi/Bose case, respectively. In the following for
the sake of clarity we will denote the Fermi and Bose field by $\psi$ and $\varphi$, respectively,
and use the letter $\phi$ for a generic field. In the case of the Dirac Field in one space
dimension, the anticommutation relations $\{\psi_n,\psi_m^\dag\}=\delta_{nm}$ and
$\{\psi_n,\psi_m\}=0$ require the field operator to be {\em non-local}. In terms of local Pauli
operators $\sigma^\alpha_n=\ldots I\otimes I\otimes\underbrace{\sigma^\alpha}_{n-\rm{th}}I\otimes
I\ldots$ the field anticommutation relations can be achieved by a Clifford-algebraic construction,
\eg $\psi_n=\left(\prod_{j=-\infty}^{n-1}\sigma_j^z\right)\sigma_n^-$. For the Bose field we can
keep the field local as long as it satisfies equal-time commutation relations of the Newton-Wigner
\cite{Newton1949p3992} form $[\varphi_n,\varphi_m^\dag]=\delta_{nm}$ and $[\varphi_n,\varphi_m]=0$.
The circuit will produce the unitary evolution of the field $\phi(t)=U_t^\dag\phi(0)U_t$, and the
unitary transformation defines the Hamiltonian $H$ through the identity
$U_t=:\exp\left(-\frac{i}{\hbar}t\hbar\omega H\right)$, where we conveniently take the Hamiltonian
as adimensional. This is equivalent to the Heisenberg-picture evolution of the field
\begin{equation}\label{Schpict}
i\hbar\partial_t\phi_n=[\phi_n,\hbar\omega H].
\end{equation}
For Hamiltonian
\begin{equation}\label{Hs}
H_s=-s\frac{i}{2}\sum_n(\phi_n^\dag\phi_{n+1}-\phi_{n+1}^\dag\phi_n)=s\frac{a}{\hbar}P,\quad
s=\pm1,\;\;P=-i\hbar\int\d x\phi^\dag(x)\partial_x\phi(x), 
\end{equation}
$P$ being the field momentum, using the identity $[AB,C]=A[B,C]_\pm\mp[A,C]_\pm B$, one obtains
$[\phi(x),H_s]=-sa^{-\frac{1}{2}}\tfrac{i}{2}(\phi_{n+1}-\phi_{n-1})=-sia\partial_x\phi(x)$ for both the
Bose and the Fermi field, where we used the identity
\begin{equation}
\partial_x\phi(x)=\frac{1}{2a}(\phi(na+a)-\phi(na-a))= \frac{1}{2a^{3/2}}(\phi_{n+1}-\phi_{n-1}).
\end{equation}
Then we can see that for $\omega a=c$ the field satisfies the massless scalar Klein-Gordon equation
$\square\phi=0$, corresponding to the two decoupled fields $i\hbar\partial_t \phi^{(s)}(x)=-
isc\hbar\partial_x\phi^{(s)}=s cp\phi^{(s)}$. In the Fermi case the two components for $s=\pm$ of
the field can be interpreted as particle and antiparticle, and the filling of the Dirac sea is
simply the reversal of the qubits (see also the following subsection on the vacuum).

The Hamiltonian $H_s$ is global, involving a whole slide of the circuit. However, using the
Trotter's formula we can see that the time evolution can be achieved with a slab of $N$ couples of
intercalated layers of bipartite gates, as in the computational circuit in Fig.
\ref{fig:synchronous}, upon writing
\begin{equation}\label{trot}
\begin{split}
U_t=e^{-i\omega t H}=\lim_{N\to\infty} U_t^{(N)},\quad U_t^{(N)}:=
\left[ \left(\prod_l
    e^{-i\frac{\pi\omega t}{4N}H_{2l-1,2l}}\right)\left(\prod_l
    e^{-i\frac{\pi\omega t}{4N}H_{2l,2l+1}}\right)\right]^N\!\!,
\end{split}
\end{equation}
for gate Hamiltonian 
\begin{equation}
H_{n,n+1}=\mp \frac{2i}{\pi}(\phi_{n+1}^\dag\phi_n-\phi_n^\dag\phi_{n+1}),
\end{equation}
where the coupling for the gate Hamiltonian has been chosen for later consistency of time intervals
with the relation $\omega a=c$. I will give a study of convergence of the limit in Eq. (\ref{trot})
elsewhere.  Here I just notice that the simple Suzuki bound for the Trotter's formula
\cite{Suzuki1976p3993} is of no use, since one would get
\begin{equation}\label{suz}
\n{U_t-U_t^{(N)}}\leq
\tfrac{\n{H_{0,1}}^2\pi^2\omega^2 t^2(2N_x+1)^2}{2N}e^{\frac{\pi\omega t}{2}
(2N_x+1)\frac{N+2}{N}\n{H_{0,1}}}, 
\end{equation}
upon considering a finite circuit with $l$ running from $-N_x$ to $+N_x$ in Eq.  (\ref{trot}).  The
bound (\ref{trot}) guarantees convergence only for $N_x$ fixed (and Fermi field, in order to have
$\n{H_{0,1}}$ bounded), namely $a$ fixed for fixed width $L$ of the circuit.  This, however, would
correspond to maximal causal velocity $\upsilon_{caus}\to\infty$, since $\tau=t/(2N)$.  In order to
keep $\upsilon_{caus}=\frac{a}{\tau}=c$ for $a$ fixed, we need to increase the width of the circuit,
so that $\frac{L}{t}=\frac{2N_x a}{2N\tau}=c$, namely $N_x=N$, but this will blow up the bound for
$N\to\infty$. Also we see that the Hamiltonian will achieve the swap for phase $\pi/2$ (modulo local
unitaries), and this corresponds to imposing $\frac{\pi\omega t}{4 N}=\frac{\pi}{2}$, namely the
time $t=N\frac{T}{\pi}$ is discrete, with $T=\frac{2\pi}{\omega}$ the oscillation period. But now, since
$\omega a=c$, one has that both grains of space and time are dictated by $\omega$, and one has
\begin{equation}
a=\frac{cT}{2\pi},\quad\tau=\frac{T}{2\pi},\quad t=2N\tau. 
\end{equation}
We remember that the angular frequency $\omega$ is only a fictitious quantity designed to simulate
the field theory, whence, it is up to us to rescale it in order to make $a$ and $\tau$ as small as
we want. Notice that $\omega$ rescales as $\omega\propto a^{-1}$, corresponding to a resulting
extensivity of $H$ versus the number of gates.

\medskip
It is obvious that we can obtain different field theories by using different realizations of the
field operator in terms of local operators, and by making different choices of the local gates.
When the Hamiltonian involves a number $2\leq k\leq\infty$ of contiguous systems, the evolution can
be achieved with a repeated slab of $k$ intercalated layers of $k$-partite gates via the Trotter
formula, corresponding to an homogeneous and isotropic circuit satisfying local causality, and thus
leading to relativistic invariance. 
 
\paragraph{Simulating QFT${}_2$ by QCFT${}_2$: the Dirac equation.}
We want now to make a quantum computer simulation of the Dirac equation, which is given by
\begin{equation}\label{Dirac}
i\hbar\partial_t\psi=
\begin{pmatrix}ic\hbar\sigma_x \partial_x & mc^2 \\ mc^2 & -ic\hbar\sigma_x \partial_x\end{pmatrix}
\psi,\qquad \psi(x)=\begin{pmatrix}\psi^1(x) \\ \psi^2(x) \\ \psi^3(x)\\ \psi^4(x)\end{pmatrix}
:=\begin{pmatrix} u(x) \\ v(x)\end{pmatrix},
\end{equation}
where
\begin{equation}
\{\psi^\alpha(x),\psi^\dag{}^\beta(y)\}=\delta_{\alpha\beta}\delta(x-y),\qquad 
\{\psi^\alpha(x),\psi^\beta(y)\}=0.
\end{equation}
In the computational representation, the field operators can be written in terms of local
single-qubit operators using the Clifford algebra as follows 
\begin{equation}\label{Clifford}
\psi_n^\alpha=\Gamma_{4n+\alpha},\quad\Gamma_k:=
\left(\prod_{j=-\infty}^{k-1}\sigma_j^z\right)\sigma_k^-,\quad \{\Gamma_k,\Gamma_h\}=\delta_{kh},
\end{equation}
where we discretize as usual as $\psi_n^\alpha=a^{\frac{1}{2}}\psi^\alpha(na)$.
Eq. (\ref{Dirac}) can be derived in the Heisenberg picture (\ref{Schpict}) from the Hamiltonian
\begin{equation}
\hbar\omega H=\int\d x\;\psi^\dag(x)\begin{pmatrix}ic\hbar\sigma_x \partial_x 
& mc^2 \\ mc^2 & -ic\hbar\sigma_x \partial_x\end{pmatrix}\psi(x)
\end{equation}

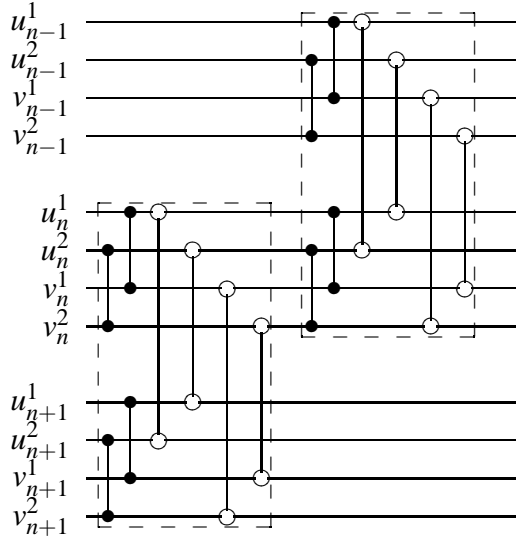
\begin{figure}
$\Qcircuit @C=.7em @R=.3em @!R {
\pureghost{{u^1_{n-1}}}&\lstick{{u^1_{n-1}}}&\qw&\qw&\qw&\qw&\qw&\qw
&\qw&\qw&\ctrl{2}&\ctrlo{6}&\qw&\qw&\qw&\qw&\qw\\
\pureghost{{u^2_{n-1}}}&\lstick{{u^2_{n-1}}}&\qw&\qw&\qw&\qw&\qw&\qw
&\qw&\ctrl{2}&\qw&\qw&\ctrlo{4}&\qw&\qw&\qw&\qw\\
\pureghost{{v^1_{n-1}}}&\lstick{{v^1_{n-1}}}&\qw&\qw&\qw&\qw&\qw&\qw&\qw
&\qw&\control\qw&\qw&\qw&\ctrlo{6}&\qw&\qw&\qw\\
\pureghost{{v^2_{n-1}}}&\lstick{{v^2_{n-1}}}&\qw&\qw&\qw&\qw&\qw&\qw&\qw
&\control\qw&\qw&\qw&\qw&\qw&\ctrlo{4}&\qw&\qw\\ 
\\
\pureghost{u^1_{n}}&\lstick{u^1_{n}}&\qw&\ctrl{2}&\ctrlo{6}&\qw&\qw&\qw&\qw
&\qw&\ctrl{2}&\qw&\controlo\qw&\qw&\qw&\qw&\qw\\
\pureghost{u^2_{n}}&\lstick{u^2_{n}}&\ctrl{2}&\qw&\qw&\ctrlo{4}&\qw&\qw&\qw
&\ctrl{2}&\qw&\controlo\qw&\qw&\qw&\qw&\qw&\qw\\
\pureghost{v^1_{n}}&\lstick{v^1_{n}}&\qw&\control\qw&\qw&\qw&\ctrlo{6}&\qw&\qw
&\qw&\control\qw&\qw&\qw&\qw&\controlo\qw&\qw&\qw\\
\pureghost{v^2_{n}}&\lstick{v^2_{n}}&\control\qw&\qw&\qw&\qw&\qw&\ctrlo{4}&\qw
&\control\qw&\qw&\qw&\qw&\controlo\qw&\qw&\qw&\qw\\
\\
\pureghost{u^1_{n+1}}&\lstick{u^1_{n+1}}&\qw&\ctrl{2}&\qw&\controlo\qw&\qw&\qw&\qw
&\qw&\qw&\qw&\qw&\qw&\qw&\qw&\qw\\
\pureghost{u^2_{n+1}}&\lstick{u^2_{n+1}}&\ctrl{2}&\qw&\controlo\qw&\qw&\qw&\qw&\qw
&\qw&\qw&\qw&\qw&\qw&\qw&\qw&\qw\\
\pureghost{v^1_{n+1}}&\lstick{v^1_{n+1}}&\qw&\control\qw&\qw&\qw&\qw&\controlo\qw&\qw
&\qw&\qw&\qw&\qw&\qw&\qw&\qw&\qw\\
\pureghost{v^2_{n+1}}&\lstick{v^2_{n+1}}&\control\qw&\qw&\qw&\qw&\controlo\qw&\qw&\qw
&\qw&\qw&\qw&\qw&\qw&\qw&\qw&\qw\gategroup{6}{3}{14}{8}{.6em}{--}
\gategroup{1}{10}{9}{15}{.6em}{--}}$
\caption{\rm Circuit for the Hamiltonian (\ref{HDirac}) for the Dirac field.}\label{f:Dirac} 
\end{figure}

Using the identity
\begin{equation}
\left[\sum_{n\alpha}\psi_n^\alpha{}^\dag K\psi_n^\alpha,\psi_l^\beta\right]
=-\sum_{n\alpha}\{\psi_n^\alpha{}^\dag,\psi_l^\beta\}K\psi_n^\alpha
=-K\psi_l^\beta,
\end{equation}
$K$ being a linear operator over the operator vector $\{\psi_n^\alpha\}$, we get the Hamiltonian
\begin{equation}\label{infevol}
H=\begin{pmatrix} \frac{i}{2}\sigma_x(\delta_+-\delta_-)  &\frac{a}{\lambda}I\\
\frac{a}{\lambda}I & -\frac{i}{2}\sigma_x(\delta_+-\delta_-)\end{pmatrix},
\end{equation}
for $\omega a=c$, and using the identities (valid at ${\cal O}(a^2)$)
\begin{equation}
\frac{1}{2a}(\delta_+-\delta_-)=\partial_x,\qquad \frac{1}{2}(\delta_++\delta_-)=1.
\end{equation}
where $\lambda:=\frac{\hbar}{mc}=3.86159*10^{-13}$ is the reduced Compton wavelength (roughly the
uncertainty in position corresponding to sufficient energy to create another particle).  The unitary
transformation can be achieved by a computational network as in Fig. \ref{fig:synchronous}, where
each wire is actually a quadruple wire as in Fig. \ref{f:Dirac}.  Here the two different types of
bipartite gates of the intercalated layers in Fig. \ref{fig:synchronous}
are represented in detail. The bonds linking the open circles represent the $\frac{i}{2}\sigma_x$
matrix blocks, whereas those linking full circles represent the $\frac{a}{\lambda}I$ blocks.

\paragraph{The vacuum.}
In our qubit description of the Dirac field the vacuum will be given by the state
$|0\>=\ldots|\downarrow\>|\downarrow\>|\downarrow\>\ldots$ and the Clifford realization of the field
in Eq. (\ref{Clifford}) will give $\psi_n|0\>=0$. The state $|\psi_n\>:=\psi_n^\dag|0\>$ will
describe a single-particle excitation, $\psi_n^\dag\psi_m^\dag|0\>$ a two-particle excitation, etc.
Notice that we could have defined the Clifford realization of the field with a $\sigma^+$ at
position $n$ for the antiparticle, and correspondingly used the state $|\uparrow\>$ in the vacuum,
defining $|0\>$ as the filled Dirac sea. An analogous representation can be used for Bosons, where
we can now have any number of particles at location $n$. It will also be handy to rewrite the
(anti)commutation relations as $[\phi_n,\phi_m^\dag]_\pm=\<\phi_n|\phi_m\>I$. Finally, it is worth
noticing that if one rewrites everything in terms of the qubit local operators there will be no role
left for the field operator (and, consequently, for the (anti)commutation relation), however, the
physics will be left untouched.

\paragraph{Simulating QFT${}_2$ by QCFT${}_2$: nonabelian gauge theories.}
Nonabelian gauge-invariance corresponds to an arbitrary choice of basis of the Hilbert space at each
wire, described by a unitary transformation $U(x)$ depending on $x$. By the no-programming theorem
\cite{niel} we know that the gauge field must be a boson, since the Hilbert space of each system
must be infinite-dimensional. Notice how in QCFT gauge invariance is natively nonabelian with the
gauge field already quantized, and already defined on a general foliation.

\paragraph{Simulating QFT${}_1$ by QCFT${}_2$.} A quantum computational circuit is in principle
capable of simulating any theory, including first-quantized ones. Which kind of computation will
simulate a first-quantized theory?  The answer is: a classical computation. In a first-quantized
field theory the field $\phi(x,t)$ is a $c$-function of position evolving in time---the so-called
``wave-function''. This will be the processed ``data'' of the computation. It will be then described
by a string $\rho_{\vec\phi}=\otimes_n|\phi_n\>\<\phi_n|$ of classical {\em infbits}
$|\phi\>\<\phi|$ with $\<\phi|\phi'\>=\delta_{\phi\phi'}$, namely orthogonal commuting projectors
corresponding to complex eigenvalues $\phi\in\Cmplx$ (the field values are discretized on a grid in
the complex plane).  This construction may look unnatural, but this is the way in which a
``second-quantized'' computational circuit will simulate a first-quantized field, as in a
Runge-Kutta integration. The Schr\H{o}dinger equation is just a special case of QFT${}_1$: this
shows how QCFT can also account for the ``quantization rule'' and the Planck constant, which are
written in the gates, and thus emerge as intrinsic features of the fabric of space-time, not as
additional axiomatic elements of QT. A general classical information processing will be described by
a classical channel, of the form
$\map{C}(\rho_{\vec\phi})=\sum_{\vec\phi'}p(\vec\phi'|\vec\phi)\rho_{\vec\phi'}$,
$p(\vec\phi'|\vec\phi)$ denoting a conditional probability.  In a deterministic evolution (like the
Schr\H{o}dinger equation) we have $p(\vec\phi'|\vec\phi)=\delta(\vec\phi'-f(\vec\phi))$, $f$ a
function, namely the processing is just the functional relation $\map{C}(\rho_{\vec\phi})=
\rho_{f(\vec\phi)}$. For a quantum deterministic evolution the function will be a linear unitary
kernel $\phi_i(t+\tau)=\sum_j U_{ij}\phi_j(t)$.

\paragraph{Simulating QFT${}_1$ by QCFT${}_1$.} There is also the possibility of considering a
``first-quantized'' kind of computational circuit. This will simulate a QFT${}_1$ more efficiently
than a QCFT${}_2$. However, we will lose the usual interpretation of quantum circuit. A QFT${}_1$
describes the evolution of a vector $\vec\phi=\transp{(\ldots,\phi_{n-1},\phi_n,\phi_{n+1},\ldots)}$
of values $\phi_n$ of the wave-function at different positions $n$. The QCFT${}_2$ simulating the
theory is constrained to keep the {\em infbits} $|\phi\>\<\phi|$ as classical. This means that the
circuit can linearly combine the eigenvalues $\phi_n$ of the projectors $|\phi_n\>\<\phi_n|$,
however, without making superpositions of the kets $|\phi_n\>$. The quantum computational circuit is
thus largely squandering the tensor-product Hilbert space (besides using infbits!), within which it
is working just as a big matrix over the vector $\vec\phi$, with the gates describing
``interactions'' between different systems actually corresponding to single matrix-elements or
matrix-blocks. As we can see, this resort to {\em substituting the tensor product with the direct
  sum, having different systems corresponding to different orthogonal states of a single quantum
  system}. The evolution of the field will be now given by $\phi(0)=U_t\phi(0)$, with
$U_t=\exp(-\tfrac{i}{\hbar}t\hbar\omega H)\phi(0)$. To a term of the form
$\sum_n\phi_n^\dag\phi_{n+l}$ in the QCFT${}_2$ Hamiltonian it will correspond the matrix
$\delta_{n,n+l}$ in the QCFT${}_1$ Hamiltonian. Thus, the scheme of the quantum circuit will look
exactly the same in the two cases, only the interpretation of the gates and wires will be different.
In QCFT${}_1$ the gauge transformation is abelian, and corresponds to an arbitrary choice of phases
of the basis of the Hilbert space giving the vector representation of the field.
\paragraph{Simulating QFT${}_1$ by QCFT${}_1$: the Schr\H{o}dinger equation.}
The QCFT${}_1$ simulation of the Schr\H{o}dinger equation for the free particle
$\partial_t\phi=i\frac{\hbar}{2m}\partial_x^2\phi$, $m$ mass of the particle, will be given by
$i\hbar\partial_t\phi_n(t)=\frac{\hbar}{2ma^2}[\phi_{n+1}(t)-2\phi_n(t)+\phi_{n-1}(t)] =\hbar\omega
H\phi_n(t)$, namely one has $\omega=\frac{\hbar}{2ma^2}$ and $H=\sum_j e_{j+1,
  j}-2e_{j,j}+e_{j,j+1}$, $e_{i,j}$ denoting the matrix with all elements equal to 0 apart from the
$i,j$-th which is equal to 1. Now the frequency $\omega$ rescales versus $a$ as $\omega\propto
a^{-2}$, contradicting the extensivity of $H$, and this is a manifestation of the fact that 
the Schr\H{o}dinger equation is not Lorentz invariant.
\paragraph{Simulating QFT${}_1$ by QCFT${}_1$: the Dirac particle.}
It is interesting to look at the QCFT${}_1$ simulation of the Dirac particle. This is given by the
same Eq.  (\ref{Dirac}), but now with $\psi$ describing a $c$-function wave. The Dirac equation is
now achieved by a QCFT${}_1$ in the Schr\H{o}dinger evolution $i\hbar\partial_t\psi=\hbar\omega
H\psi$ from the Hamiltonian
\begin{equation}\label{HDirac}{\footnotesize
H=\begin{pmatrix}\ldots &\ldots &\ldots &\ldots &\ldots &\ldots &\ldots &\ldots\\
\ldots &0&0&-\frac{i}{2}\sigma_x&0&0&0&\ldots\\
\ldots &0&0 & \frac{a}{\lambda}I &\frac{i}{2}\sigma_x &0&0&\ldots\\
\ldots &\frac{i}{2}\sigma_x&\frac{a}{\lambda}I &0 & 0 & -\frac{i}{2}\sigma_x&0&\ldots\\
\ldots &0&-\frac{i}{2}\sigma_x & 0 & 0 & \frac{a}{\lambda}I&\frac{i}{2}\sigma_x&\ldots\\
\ldots &0&0 & \frac{i}{2}\sigma_x & \frac{a}{\lambda}I &0&0&\ldots\\
\ldots &0&0&0 & -\frac{i}{2}\sigma_x & \frac{a}{\lambda}I &0&\ldots\\
\ldots &\ldots &\ldots &\ldots &\ldots &\ldots &\ldots &\ldots\\ \end{pmatrix}}.
\end{equation}
The unitary transformation is achieved by the same computational network of the Dirac QCFT${}_2$,
but now the gates are substituted by the 4$\times$4 submatrices (2$\times$2 with matrix elements
made of Pauli operators) in the Hamiltonian (\ref{HDirac}).

\paragraph{Zitterbewegung.}
The zigzag motion within gates on the circuit as in Fig.  \ref{fig:synchronous} corresponds exactly
to the mysterious {\em Zitterbewegung} motion (German for "trembling motion") of the Dirac particle,
which is the oscillation at the speed of light with amplitude $\lambda$ of the position of the
particle around the median, with a circular frequency of $2 mc^2/\hbar\simeq 1.6 * 10^{21}$ Hz, and
resulting from the interference between positive and negative energy solutions.\footnote{Such motion
  was sometimes interpreted as an interaction of the classical particle with the zero-point field.
  Schr\H{o}dinger proposed the electron spin to be a consequence of the Zitterbewegung.} Notice that
in the context of QCFT, as we have see in Fig. 1, the phenomenon of the Zitterbewegung becomes a
dominant feature valid for all field theories, the frequency of the ziz-zag being an increasing
function of the particle mass.
\paragraph{The Planck constant and the quantization rule.}
From the form of the circuit Hamiltonian (\ref{HDirac}) we see that the space scale at which we can
differentiate the field must be $x\gg\frac{\hbar}{mc}$, namely above the Compton wavelength, whence
$\hbar$ defines the scale at which we can see the gates in terms of the Zitterbewegung frequency
(function of the mass) and the causal speed $\upsilon_{caus}=c$. In the same fashion we could
imagine the quantization rule for momentum $p=-i\hbar\partial_x$ as emergent from the circuit
description, with the momentum describing the ``swappiness'' of the field, namely the tendency of
the gates to swap the particle.

\subsection{The Quantum-Computational Field Theory program.}
We have seen how QCFT can in principle unify the whole field theory in a causal computational
network described only by QT.  Space and time emerge from local causality already endowed with
relativistic invariance and gauge invariance. The causal-network makes the fabric of space-time,
solving the logical problems related to either the action-at-contact or the action-at-distance
\cite{Lange2006} description of the field, and motivating the lattice structure based solely with
the causality principle and the logical separation between cause and effect. The Zitterbewegung of
the Dirac particle becomes a general phenomenon within QCFT, with the zig-zag frequency related to
the particle mass, whereas the Planck constant becomes the blurring scale of the causal network, via
the Compton wavelength of the particle. As for any lattice theory, the mathematical problems related
to the continuum (\eg ultraviolet divergences) disappear.  QCFT also provides a unified framework
for different fields, and even the quantization rule and $\hbar$ become emergent features. A point
that, due to limited space, I didn't have the chance of discussing here is that QCFT also provides a
systematic way for consistently generalizing the whole theoretical framework in view of Quantum
Gravity, by changing the computational engine from QT toward an input-output computation with no
pre-established causal relations \cite{upcomingPaolo}.

Which are the first steps to be taken in the QCFT research program? Here, I just mention the most
urgent. As a general priority, we want to rederive the Feynman's path integral via the Trotter
formula, similarly to what made in Ref.  \cite{Nakamura1991p3989}. We need a reliable bound for the
Trotter formula which is tighter than the Suzuki's.  Then, we want to understand the motivation of
the nonlocal operator realization of fields, which are necessary for the (anti)commutation
relations, avoiding Grassman variables, and recover the quantization rules as emergent from the
computational network.  The full problem of the microcausality conditions and spin statistics and
para-statistics \cite{Teller1995} needs to be re-addressed within QCFT. We want to derive a toy
natively nonabelian gauge theory of a Dirac particle with e.m. field, and explore the connections
between QCFT${}_2$ and existing lattice theories.

In taking the long route of the QCFT program, a strong motivation to persevere is to remind us of
how many problems plague QFT. From the QCFT program we will definitely learn more about the
foundations of QFT.

\subsection*{Acknowledgments.} 
I thank Masanao Ozawa, Paolo Perinotti, Arkady Plotnitski, Philip Goyal, Lucien Hardy, Rafael
Sorkin, and Lee Smolin for very interesting discussions and suggestions, and for their support and
encouragement. In particular, I thank Masanao Ozawa for pointing me out Ref.
\cite{Nakamura1991p3989}, and Paolo Perinotti for reminding me the Zitterbewegung phenomenon.

\end{document}